\documentclass[aps,prd,preprint,showpacs]{revtex4}
\usepackage{latexsym}
\usepackage{amssymb,amsmath}
\usepackage{graphicx}
\begin{document}

\thispagestyle{empty}

{\baselineskip0pt
\leftline{\large\baselineskip16pt\sl\vbox to0pt{\hbox{\it Department of
Mathematics and Physics}
               \hbox{\it Osaka City  University}\vss}}
\rightline{\large\baselineskip16pt\rm\vbox to20pt{\hbox{OCU-PHYS-317}
            \hbox{AP-GR-69}
\vss}}%
}

\vskip0.5cm

\title{Maximal slicing of $D$-dimensional spherically-symmetric
\\
 vacuum spacetime}
\author{$^1$Ken-ichi Nakao\footnote{knakao@sci.osaka-cu.ac.jp}, 
$^1$Hiroyuki Abe\footnote{abe@sci.osaka-cu.ac.jp}
$^2$Hirotaka Yoshino\footnote{hyoshino@phys.ualberta.ca} 
and $^3$Masaru Shibata\footnote{shibata@yukawa.kyoto-u.ac.jp}
}
\affiliation{$^1$Department of Mathematics and Physics, Graduate School of Science, 
Osaka City University, Osaka 558-8585, Japan\\
$^2$Department of Physics, University of Alberta, Edmonton, Alberta, Canada 
T6G 2G7 \\
$^3$Yukawa Institute for Theoretical Physics, Kyoto University, 
Kyoto 606-8502,~Japan
}
\date{\today}

\begin{abstract}                
We study the foliation of a $D$-dimensional spherically symmetric
black-hole spacetime with $D\ge 5$ by two kinds of one-parameter
family of maximal hypersurfaces: a reflection-symmetric foliation with 
respect to the wormhole slot and a stationary foliation that has an
infinitely long trumpet-like shape.  As in the four-dimensional case, 
the foliations by the maximal hypersurfaces have the singularity
avoidance nature irrespective of dimensionality.  This indicates that
the maximal slicing condition will be useful for simulating
higher-dimensional black-hole spacetimes in numerical relativity.  For
the case of $D=5$, we present analytic solutions of the intrinsic 
metric, the extrinsic curvature, the lapse function, and the shift
vector for the foliation by the stationary maximal hypersurfaces.  This 
data will be useful for checking five-dimensional numerical relativity 
codes based on the moving puncture approach. 
\end{abstract}


\maketitle

\vskip1cm
\section{Introduction}

The merger of binary black holes is one of the most important sources
of gravitational waves for gravitational-wave detectors. As a result
of longterm effort by many numerical-relativity researchers, it is now
feasible to theoretically predict the orbital evolution and merger
process of astrophysical binary black holes and resulting
gravitational waves emitted from these systems by numerical
relativity.  After the first report of longterm simulations of binary
black holes by Pretorius~\cite{Pretrius}, several groups have also
succeeded in longterm
simulations~\cite{Campanelli,Baker,Herrmann,Sperhake,Scheel,Bergmann}.
There are basically three formulations in numerical relativity,
workable for simulating dynamical black-hole system; one is the
``generalized harmonic'' formulation with the help of ``black hole
excision''~\cite{Pretrius}, second one is the so-called
Baumgarte-Shapiro-Shibata-Nakaumra (BSSN) 
formulation~\cite{SN1995,BS1999} together with the ``moving puncture''
method, and third one is the hyperbolic formulation with the black
hole excision \cite{Scheel}.

Another scenario of black hole formation in the system of two
relativistic objects was pointed out in the non-astrophysical context
\cite{BF99,Dimo,Giddings}: mini-black-hole production at high-energy
particle collisions in particle accelerators in the framework of the
brane world scenarios (i.e., the so-called TeV gravity scenarios)
\cite{Arkani,RS99}.  Motivated by this possibility, mini black holes
in the particle colliders have been studied by many researchers in the
past decade from a wide viewpoints (see \cite{review} for a recent
review).  There are two marked differences between the black hole
production in the colliders and the black hole mergers in
astrophysical situations.  One is the relative speed of the objects.
In the high-energy particle collisions at the LHC, the $\gamma$-factor
of an incoming proton is $\sim 7\times 10^{3}$, and that of the
partons consisting of each proton can be much larger.  The other is
the number $D$ of the spacetime dimensions, because the spacetime
dimensions higher than four is essential for the TeV gravity
scenarios.

In order to understand well the black hole production in the particle
colliders, numerical relativity is probably the unique approach.  To
perform a simulation of this system, two techniques have to be
developed: One is the technique to handle the high-energy objects and
the other to simulate higher-dimensional black-hole spacetimes.  The
first technique has been developed for the four-dimensional
case~\cite{Sperhake:2008ga,Shibata:2008rq,NEW} by modeling the
high-energy two-particle system as the high-velocity two-black-hole
collision and extrapolating the results to the ultra-relativistic
regime.  Such simulations have not been performed for
higher-dimensional spacetimes yet (but see Refs.~\cite{YN03,YR05}
for studies on the apparent horizon in the system of two high-energy
particles in higher dimensions). Also a numerical code to simulate
higher-dimensional spacetimes with the BSSN formalism has been recently 
developed \cite{YS09} (see a list of proposed several test simulations 
and the successful results). 

In this paper, we investigate the maximal slicing condition for the
$D$-dimensional spherically symmetric black-hole spacetime ($D\geq5$),
known as the Schwarzschild-Tangherlini solution~\cite{Tangherlini}.
This study is motivated by numerical relativity performed with the
BSSN formalism and the puncture method, by which whole spacetime
region including the black hole interior is simulated, and thus, the
slicing condition that has the singularity avoidance nature has to be
adopted.  For the four-dimensional Schwarzschild black hole, it was
shown that the sequence of maximal slices never plunges into the
curvature singularity but asymptotes to the so-called limit surface
\cite{Estabrook_etal}.  We shall show that the sequence of the maximal
slices avoids the singularity also in the higher-dimensional
Schwarzschild-Tangherlini spacetime. This gives the theoretical
foundation that both the maximal slicing condition and the puncture
gauge condition (see \cite{Alcubierre:2002kk,Hannam_etal_1}) have the
desired features in higher-dimensional numerical relativity.  Another
important aspect of this study is to derive a stationary sequence of
the maximal slicing hypersurfaces that provides a useful analytic
solution for a benchmark test of higher-dimensional numerical
relativity codes, as discussed in 
Refs.~\cite{Hannam_etal_1,Baumgarte_N} in the four-dimensional case.

This paper is organized as follows. In Sec.~II, we derive the general
solution in the maximal slicing for the $D$-dimensional spherically
symmetric black-hole spacetime.  Then, in Sec.~III, we investigate the 
foliation by the maximal hypersurfaces which are reflection symmetric
at the wormhole slot.  In Sec.~IV, we study the foliation by the
stationary maximal hypersurfaces. After explicitly describing the
analytic solution in the case $D=5$, we show its usefulness as a
benchmark test for numerical relativity, by performing a numerical
simulation adopting the analytic solution as the initial data.  Sec.~V
is devoted to a summary.  In Appendix~A, the Kruskal extension of the
Schwarzschild-Tangherlini spacetime is analyzed. 

We adopt the geometrized units $c=G=1$ throughout this paper, where $c$
is the speed of light and $G$ is the gravitational constant of a
$D$-dimensional spacetime. The Greek indices ($\mu, \nu, ...$)
represent the components of a spacetime, while the Latin indices
($i, j, ...$) represent the components of a space.

\section{General spherically symmetric maximal slicing}

Following Estabrook et al.~\cite{Estabrook_etal}, we derive general
spherically-symmetric maximal slicing of the Schwarzschild-Tangherlini
spacetime, i.e., the $D$-dimensional spherically-symmetric vacuum
black-hole solution with $D\ge 5$.  The general form of its line
element is 
\begin{equation}
ds^2=-\alpha^2 dt^2+\gamma (dr+\gamma^{-1}\beta dt)^2+r^2d\Omega_{D-2}^2,
\label{eq:metric}
\end{equation}
where $d\Omega_{D-2}^2$ is the line element of a ($D-2$)-dimensional 
unit sphere with the $(D-2)$-area 
$\Omega_{D-2}:=(D-1)\pi^{(D-1)/2}/\Gamma(\frac{D+1}{2})$, and 
$\alpha$, $\beta$, and $\gamma$ are functions of $t$ and $r$. 
 
Einstein's equations with appropriate coordinate conditions lead to
the basic equations for $\alpha$, $\beta$ and $\gamma$.  Because the
$(D-2)$-dimensional spherical-polar coordinate system is uniquely
determined in the spherically symmetric spacetime, two conditions are
required to fix the remaining two coordinates; the condition to
determine the foliation of the spacetime by a one-parameter family of
spacelike hypersurfaces and that to specify the time evolution of the
radial coordinate.  In Eq.~\eqref{eq:metric}, the radial coordinate
has already been fixed so that $\Omega_{D-2}r^{D-2}$ becomes the area
of the $(D-2)$-dimensional sphere labeled by $r$.  Thus, we only need
the condition for the foliation.  As mentioned earlier, we shall
consider the foliation of this spacetime by the family of maximal
hypersurfaces, where a maximal hypersurface implies that the trace of
the extrinsic curvature vanishes on it, i.e. $K=0$.


In the following, we solve Einstein's equations for the metric 
\eqref{eq:metric} in the maximal slicing condition.
From the maximal slicing condition $K=0$, we have 
\begin{equation}
-\partial_t\left(\ln\gamma\right)+\beta \gamma^{-1}\partial_r
\left[\ln\left(\beta^2\gamma^{-1}r^{2(D-2)}\right)\right]=0.
\label{Maximal}
\end{equation}
The Hamiltonian and momentum constraints are calculated to give 
\begin{equation}
(D-1)\alpha^{-2}\beta^2\gamma^{-1}=(D-3)(\gamma-1)+r\partial_r(\ln\gamma),
\label{HC}
\end{equation}
\begin{equation}
\partial_r\left[\ln\left(\alpha^{-1}\beta\gamma^{-1}r^{D-2}\right)\right]=0.
\label{MC}
\end{equation}
Using $\partial_t K=0$, the evolution equation for the trace 
part of the extrinsic curvature reduces to 
\begin{eqnarray}
\partial_r^2\alpha+(D-2)r^{-1}\partial_r\alpha-\frac{1}{2}
\left[\partial_r(\ln\gamma)\right]\partial_r\alpha
=(D-2)r^{-2}\left[(D-3)(\gamma-1)+r\partial_r(\ln\gamma)\right].
\label{Lapse_eq}
\end{eqnarray}
The remaining nontrivial component of the evolution equations gives 
\begin{multline}
\partial_t\left[\ln\left(\alpha^{-1}\beta\right)\right]=
\left[(2D-5)\beta\gamma^{-1}+(D-3)\alpha^2\beta^{-1}(\gamma-1)\right]r^{-1}
+3\gamma^{-1}\partial_r\beta \\
+\frac{1}{2}\left(\alpha^2\beta^{-1}-4\beta\gamma^{-1}\right)
\partial_r(\ln \gamma)
-(\beta\gamma^{-1}+\alpha^2\beta^{-1})\partial_r(\ln\alpha). 
\label{Evol_eq}
\end{multline}

General solutions for $\alpha$, $\beta$ and $\gamma$ are derived 
as follows. From Eq.~\eqref{MC}, $\beta$ is determined as 
\begin{equation}
\beta=T(t)\alpha\gamma r^{-(D-2)}, \label{beta}
\end{equation}
where $T(t)$ is a function of integration. 
Substituting the above equation into Eq.~\eqref{HC}, equation 
for $\gamma$ is derived to give 
\begin{equation}
\partial_r(r^{D-3}\gamma^{-1})=(D-3)r^{D-4}-(D-1)T^2r^{-D},
\end{equation}
and by the integration of this equation, we obtain 
\begin{equation}
\gamma^{-1}=1-\left[r_{\rm g}(t)/r\right]^{D-3}+T^2/r^{2(D-2)},
\label{gamma}
\end{equation}
where $r_{\rm g}(t)$ is a function of integration. 
Eliminating $\beta$ in Eq.~\eqref{Maximal} by using Eq.~\eqref{beta}, 
and then using Eq.~\eqref{gamma}, we have 
\begin{equation}
\partial_r(\alpha\gamma^{1/2})
=\gamma^{3/2}\left[\frac{\partial_t(r_{\rm g}^{D-3})r}{2T}
-\frac{\partial_tT}{r^{D-2}}\right].
\label{lapse}
\end{equation}
Eliminating $\beta$ from Eq.~\eqref{Evol_eq} and rewriting
the result with help of Eq.~\eqref{lapse}, we find that $r_{\rm g}$ 
is a constant: 
\begin{equation}
\partial_t r_{\rm g}=0.
\end{equation}
Using this fact, Eq.~\eqref{lapse} is integrated to give
\begin{equation}
\alpha=f(r_{\rm g}/r;T)^{1/2} 
\left[1+\frac{\partial_tT}{r_{\rm g}^{D-3}}
\int_0^{r_{\rm g}/r}x^{D-4}f(x;T)^{-3/2}dx
\right].
\label{lapse_sol}
\end{equation}
where
\begin{equation}
f(x;T):=1-x^{D-3}+T^{2}r_{\rm g}^{-2(D-2)}x^{2(D-2)}. 
\end{equation}

If $T(t)$ is determined, $\gamma$ and $\alpha$ are subsequently
derived by solving Eqs.~\eqref{gamma} and \eqref{lapse_sol}, and then,
$\beta$ is determined by Eq.~\eqref{beta}.  Because $T(t)$ is an 
arbitrary function, we have to impose an additional condition for
$T(t)$ for specifying a solution.  In the next section, we impose the
reflection symmetry with respect to the wormhole slot and derive the
function $T(t)$ that specifies this slicing.  In Sec.~IV, we also
consider the case that $T(t)$ is constant and give a different
class of the maximal hypersurfaces. 

\section{Reflection Symmetric Foliation}

The coordinate system with the choice $T=0$ agrees with 
the Schwarzschild-Tangherlini static coordinates, 
\begin{equation}
ds^2=-\left[1-\left(\frac{r_{\rm g}}{r}\right)^{D-3}\right]d\tau^2
+\left[1-\left(\frac{r_{\rm g}}{r}\right)^{D-3}\right]^{-1}dr^2
+r^2d\Omega_{D-2}^2. \label{eq:standard}
\end{equation}
To derive the coordinates for the general foliation with $T\neq 0$, we
first prepare the maximally extended Schwarzschild-Tangherlini
spacetime (see Appendix~A for the Kruskal extension), and then, 
perform coordinate transformation from the
Schwarzschild-Tangherlini coordinates $(\tau,r)$ to the coordinates
$(t,r)$ for a foliation $T\neq 0$.  Here, the
Schwarzschild-Tangherlini time coordinate $\tau$ is given as
$\tau=\tau(t,r)$ and satisfies 
\begin{eqnarray}
\frac{\partial \tau}{\partial t}
&=&\alpha \gamma^{1/2}, 
\label{eq:dtaudt}\\
\frac{\partial \tau}{\partial r}
&=&
-\gamma^{1/2}Tr^{-(D-2)}\left[1-\left(\frac{r_{\rm g}}{r}\right)^{D-3}\right]^{-1}.
\label{eq:dtaudr}
\end{eqnarray}
Integrating Eq.~\eqref{eq:dtaudr} gives $\tau$ in the form 
\begin{equation}
\tau=Tr_{\rm g}^{-D+3}\int_{r_{\rm g}/r}^{X(T)}
\frac{x^{D-4}}{(x^{D-3}-1)
f(x;T)^{1/2}}dx,
\label{tau-integral_form}
\end{equation}
where $X(T)$ is a function of integration. Substituting the above equation into 
Eq.~\eqref{eq:dtaudt}, the equation for $X$ is derived as 
\begin{equation}
\frac{dX}{dT}=T^{-1}X^{4-D}\left(X^{D-3}-1\right) f(X;T)^{1/2} 
\left[
\frac{r_{\rm g}^{D-3}}{\partial_tT} +\int_0^X 
\frac{x^{D-4}}{f(x;T)^{3/2}}dx
\right]. \label{eq:X-eq}
\end{equation}

\begin{figure}
\centering
\includegraphics[width=0.45\textwidth]{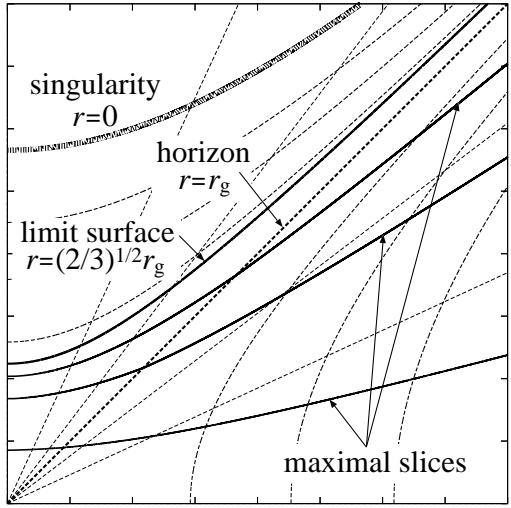}
\caption{
The reflection symmetric maximal slicing in the Kruskal diagram 
of Schwarzschild-Tangherlini spacetime with $D=5$. 
The maximal hypersurfaces with the reflection symmetry 
with respect to the wormhole slot $\tau=0$ are depicted by solid curves.
The dotted curves show the $r=\textrm{const.}$ lines, while
the dotted straight lines indicate the $\tau=\textrm{const.}$ lines.
The singularity, the limit surface, and the horizon are 
given by $r=0$, $\sqrt{2/3}r_{\rm g}$, and $r_{\rm g}$, respectively. 
}
\label{fg:symmetric}
\end{figure}

Here, we require the hypersurface to have a reflection symmetry with
respect to the wormhole slot, i.e., $\tau=0$ , which is located in 
the black hole interior $r=r_{\rm min} < r_{\rm g}$ (see the Kruskal diagram
Fig.~\ref{fg:symmetric}); this condition determines $T(t)$.  At
$\tau=0$, the 1-form $\nabla_\mu t$, normal to the hypersurface labeled
by $t$, should be perpendicular to $(\partial/\partial\tau)^\mu$ because
$\tau$ is a spacelike coordinate in the black hole interior, and thus
\begin{equation}
\left(\frac{\partial}{\partial \tau}\right)^\mu\nabla_\mu t
=\frac{1}{\alpha\gamma^{1/2}}=0 \qquad
{\rm at} \quad \tau=0.
\end{equation}
Because $\alpha$ should be finite everywhere on the slice, 
the following condition has to be satisfied: 
\begin{equation}
\frac{1}{\gamma(r_{\rm min};T)}=0.
\label{eq:rmin}
\end{equation}
Note that the circumferential radius $r$ takes the minimal value $r_{\rm min}$ at 
$\tau=0$ in the black hole interior because of
the requirement of the reflection symmetry.  Equation (\ref{eq:rmin})
determines the value of $r_{\rm min}$ for a given value of $T$,
i.e., $r_{\rm min}=r_{\rm min}(T)$.

The equation $f(x;T)=0$ has at most two real positive roots, and the
smaller one is $x=r_{\rm g}/r_{\rm min}$ because $f(r_{\rm
g}/r;T)=1/\gamma(r;T)$ holds.  By a careful limiting procedure, the
following fact is found: If $X(T)$ is a smaller root of $f(X;T)=0$,
then $X$ satisfies Eq.~\eqref{eq:X-eq}; and if $X$ is a larger root,
then the integrand in Eq.~\eqref{eq:X-eq} becomes imaginary that is
forbidden here. This implies that we have to adopt $X(T)=r_{\rm g}/r_{\rm
min}(T)$ in Eq.~\eqref{tau-integral_form}. 

We further require the time coordinate $t$ to agree with 
the Schwarzschild-Tangherlini time coordinate 
$\tau$ at spacelike infinity $r\rightarrow\infty$. Then, 
Eq.~\eqref{tau-integral_form} gives 
\begin{equation}
t=Tr_{\rm g}^{-D+3}\int_0^{X(T)}
\frac{x^{D-4}}{(x^{D-3}-1)
\sqrt{f(x;T)}}dx. 
\label{eq:T-sol}
\end{equation}
This equation determines the function of $T(t)$. 

If the equation $f(x,T)=0$ has a degenerate double root, the integral
of Eq.~\eqref{eq:T-sol} diverges, i.e., $t=\infty$. This is the
so-called limit surface to which the sequence of maximal hypersurfaces
asymptotes.  Since the root of $f=0$ is also the root of the equation
$df/dx=0$ in this case, we obtain 
\begin{equation}
\lim_{t\rightarrow\infty}T^2(t)=T_\infty{}^2:=\frac{D-3}{2(D-2)}
\left[\frac{D-1}{2(D-2)}\right]^{{(D-1)}/{(D-3)}}r_{\rm g}{}^{2(D-2)}.
\label{def:T_infinity}
\end{equation}
The root $x=x_{\rm lim}$ of $f(x,T_\infty)=0$ is then given by
\begin{equation}
x_{\rm lim}=\left[\frac{2(D-2)}{D-1}\right]^{{1}/{(D-3)}}.
\label{formula:xlim}
\end{equation}
The minimal radius $r_{\rm min}$ for $t\rightarrow\infty$ is called
the limit radius, and it is given by
\begin{equation}
r_{\rm lim}=\frac{r_{\rm g}}{x_{\rm lim}}
=\left[\frac{D-1}{2(D-2)}\right]^{{1}/{(D-3)}}r_{\rm g}.
\label{formula:rlim}
\end{equation}
In the case of $D=4$, this shows the known result $r_{\rm lim}=(3/4)r_{\rm
g}$ \cite{Estabrook_etal}. 

Several maximal hypersurfaces with the reflection symmetry 
at the wormhole slot $\tau=0$ 
for the case of $D=5$ are depicted in the Kruskal diagram 
Fig.~\ref{fg:symmetric} (see Appendix A for the method of
embedding). As in the case $D=4$, the sequence of maximal hypersurfaces
has the singularity avoidance nature for $D\ge 5$.

\section{Stationary Foliation}

In this section, we turn our attention to a foliation of maximal
hypersurfaces which is different from the one analyzed in Sec.~III;
foliations for which $T(t)$ is constant and the reflection symmetry
with respect to the wormhole slot is not imposed in general.  As the
fixed value, we choose $T=T_{\infty}$ defined in 
Eq.~\eqref{def:T_infinity}.  In this case, the sequence of
hypersurfaces is one-parameter family labeled by $t$ (not by $T$),
and the metric does not depend on the time coordinate $t$.  Because
the time coordinate basis is not orthogonal to each hypersurface, we
refer to this foliation as the stationary foliation.

\begin{figure}
\centering
\includegraphics[width=0.45\textwidth]{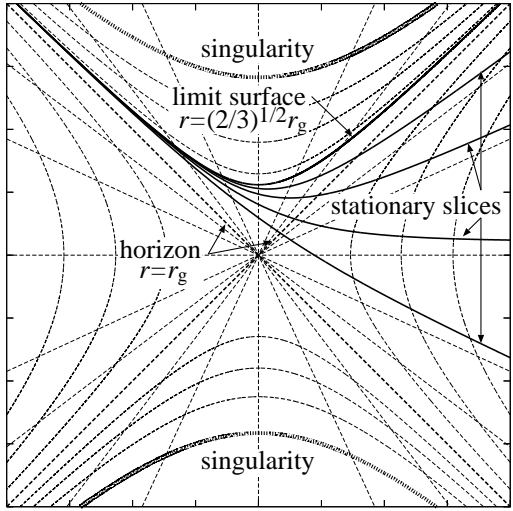}
\caption{
The stationary maximal slicings of $T=T_\infty$ 
in the Kruskal diagram of Schwarzschild-Tangherlini spacetime with $D=5$
(solid lines). Except for the limit surface $r=\sqrt{2/3}r_{\rm g}$,
the maximal hypersuefaces of $T=T_\infty$ are
not symmetric with respect to the wormhole slot. 
}
\label{fg:stationary}
\end{figure}


Several stationary maximal hypersurfaces of $T=T_\infty$ for $D=5$ are
depicted in the Kruskal diagram Fig.~\ref{fg:stationary} (see Appendix
A for the method of embedding). The limit surface exists also for the
stationary foliation with $T=T_{\infty}$, and it agrees with the limit
surface of the reflection-symmetric foliation, studied in the previous
section.  Figure~\ref{fg:stationary} shows that the sequence of
maximal hypersurfaces in this class also has the singularity
avoidance nature and that except for the limit surface, the
hypersurfaces of the stationary foliation are not reflection 
symmetric. 

The stationary foliation is of special interest in connection with
numerical relativity performed in the moving puncture approach,
because it would be an attractor of the time evolution with a
dynamical slicing ($\partial_t \alpha=-C\alpha K$ where $C$ is a
constant) \cite{Bona_etal} and $\Gamma$-driver conditions
\cite{Alcubierre:2002kk} as demonstrated in \cite{Hannam_etal_1} for
$D=4$ and in \cite{YS09} for $D=5$. In other words, the numerical
evolution starting from a hypersurface of stationary foliation has
to be unchanged in time in these gauge conditions. Therefore, this
solution provides a useful benchmark test for higher-dimensional
numerical relativity codes. 

Black hole simulation in numerical relativity is often performed in
the isotropic coordinates.  In Sec. IVA, we describe hypersurfaces of
stationary foliation in the isotropic coordinates, and study the
asymptotic behaviors of the spatial metric.  In Sec. IVB, we derive
the analytic solution of the stationary maximal hypersurface for $D=5$
in terms of the BSSN variables, and then, demonstrate its usefulness
for a benchmark test of numerical relativity codes, by performing
numerical simulation. 

\subsection{Asymptotic behaviors}

In the isotropic coordinates, the line element of a
$(D-1)$-dimensional spherically-symmetric spacelike hypersurface is
written as
\begin{equation}
dl^2=\psi^{4/(D-3)}\left(dR^2+R^2d\Omega_{D-2}\right).
\label{eq:metric_isotropic}
\end{equation}
By comparing this with the spatial part of the metric~\eqref{eq:metric},
the relation between the coordinates $r$ and $R$ is found: 
\begin{eqnarray}
\gamma^{1/2}dr&=&\psi^{2/(D-3)}dR, \\
r&=&\psi^{2/(D-3)}R. \label{eq:psi}
\end{eqnarray}
These equations lead to a differential equation
\begin{equation}
\frac{d\ln R}{d\ln r}=\frac{1}{\sqrt{f(r_{\rm g}/r; T_{\infty})}},
\end{equation}
and the formal solution is given by
\begin{equation}
R=R_{\rm c}e^{I(r)},
\label{formal_solution_R}
\end{equation}
where
\begin{equation}
I(r)=\int_{r_{\rm g}/r}^1\frac{dx}
{x \sqrt{f(x;T_{\infty})}}. 
\label{eq:I-def}
\end{equation}
In the following, we analyze asymptotic relations between $r$ and $R$, 
and the behavior of the conformal factor $\psi$
in the distant region $r\gg r_{\rm g}$ 
and in the neighborhood of $r=r_{\rm lim}$
(given by Eq.~\eqref{formula:rlim}), one by one. 

In order to study the asymptotic behavior of $R$ in the distant region, 
we rewrite the function $I(r)$ in the form
\begin{equation}
I(r)=-\ln\left({r_{\rm g}}/{r}\right)+F(r_{\rm g}/r),
\end{equation}
where
\begin{equation}
F(y):=\int_y^1dx\frac{1-\sqrt{f(x;T_{\infty})}}
{x\sqrt{f(x;T_{\infty})}}.
\end{equation} 
For analyzing its behavior for $y \rightarrow 0$, we 
write $F(y)$ in the form of a Maclaurin series, 
\begin{equation}
F(y)=F(0)-\frac{y^{D-3}}{2(D-3)}+{\cal O}(y^{2D-7}). 
\end{equation}
By requiring $R$ to agree with $r$ at spatial infinity $r\to \infty$,
the integration constant $R_{\rm c}$ is chosen to be
\begin{equation}
R_{\rm c}=r_{\rm g}e^{-F(0)}.
\end{equation}
Then, the relation between $R$ and $r$ is found to be 
\begin{equation}
R\simeq r\left[1-\frac{1}{2(D-3)}\left(\frac{r_{\rm g}}{r}\right)^{D-3}\right]
\qquad {\rm for} \quad r\gg r_{\rm g}. 
\label{eq:R-asymp}
\end{equation}
From Eqs.~\eqref{eq:psi} and \eqref{eq:R-asymp},  
the asymptotic behavior of the conformal factor $\psi$ is given by
\begin{equation}
\psi\simeq 1+\frac{1}{4}\left(\frac{r_{\rm g}}{R}\right)^{D-3} 
\qquad {\rm for} \quad R\gg r_{\rm g}, 
\end{equation}
which is the well-known relation for $D=4$.

Next, we investigate the asymptotic behavior of $R$ in the
neighborhood of $r=r_{\rm lim}$. 
Because $f(x;T_{\infty})=0$ has the degenerate double root at 
$x=x_{\rm lim}$ given by Eq.~\eqref{formula:xlim}, 
$f(x;T_{\infty})$ is written in the form 
\begin{equation}
f(x;T_\infty)=(x_{\rm lim}-x)^2 h(x). 
\label{eq:factorize}
\end{equation}
Here, a function $h(x)$ is a positive definite polynomial of $x$ for
$0<x\leq x_{\rm lim}$.  Substituting Eq.~(\ref{eq:factorize}) 
into Eq.~(\ref{eq:I-def}), the function $I(r)$ is rewritten in the 
following form, 
\begin{eqnarray}
I(r)
:=\int_{r_{\rm g}/r}^1\frac{dx}{x(x_{\rm lim}-x)\sqrt{h(x)}}
=\frac{1}{x_{\rm lim}\sqrt{h(x_{\rm lim})}}
\ln\left|\frac{r-r_{\rm lim}}{r(1-r_{\rm lim}/r_{\rm g})}\right|
+H(r_{\rm g}/r),
\label{integral_I_origin}
\end{eqnarray}
where
\begin{equation}
H(y):=\frac{1}{x_{\rm lim}\sqrt{h(x_{\rm lim})}}
\int_y^1dx\frac{x_{\rm lim}\sqrt{h(x_{\rm lim})}-x\sqrt{h(x)}}
{x(x_{\rm lim}-x)\sqrt{h(x)}}.
\end{equation}
Note that $H(y)$ is finite at $y=x_{\rm lim}$. 
By taking the second derivative of Eq.~\eqref{eq:factorize}, we have 
\begin{equation}
h(x_{\rm lim})=\frac{1}{2}\frac{d^2f}{dx^2}\biggr|_{x=x_{\rm lim}}
=\frac{1}{2}{(D-1)(D-3)}\left[\frac{2(D-2)}{D-1}\right]^{(D-5)/(D-3)}.
\end{equation}
Substituting this expression with Eq.~\eqref{formula:xlim} 
into Eq.~\eqref{integral_I_origin} and using Eq.~\eqref{formal_solution_R},
we obtain the relation between $R$ and $r$ in the neighborhood of $r= r_{\rm lim}$ 
\begin{equation}
R \simeq R_0\left(\frac{r}{r_{\rm lim}}-1\right)^{1/\sqrt{(D-3)(D-2)}},
\end{equation}
where
\begin{equation}
R_0=R_ce^{H(x_{\rm lim})}
\left(1-\frac{r_{\rm lim}}{r_{\rm g}}\right)^{-1/\sqrt{(D-3)(D-2)}}.
\end{equation}
Thus, the asymptotic behavior of the conformal factor is given by
\begin{equation}
\psi\simeq \left(\frac{r_{\rm lim}}{R}\right)^{(D-3)/2} \qquad {\rm for} \quad
R \ll r_{\rm g}.
\end{equation}
Although the factor $\psi$ becomes steeper near the puncture for higher
dimensions, the overall conformal factor $\psi^{4/(D-3)}$ in 
Eq.~\eqref{eq:metric_isotropic} has the universal behavior
$\psi^{4/(D-3)}\simeq (r_{\rm lim}/R)^2$ for arbitrary values of $D$.

The above result shows that the coordinate origin $R=0$ has a finite
circumferential radius $r=r_{\rm lim}$. By contrast, the proper
length $l$ from a point labeled with the isotropic radial coordinate
$R$ to the origin $R=0$ is 
\begin{equation}
l=\int_0^R\psi^{2/(D-3)}(R^\prime)dR^\prime 
\sim 
r_{\rm lim} \int_0^R\frac{dR^\prime}{R^\prime}
=\infty.
\end{equation}
This result implies that each stationary maximal hypersurface with
$T=T_\infty$ has an infinitely long trumpet shape for $D\geq4$. 

\subsection{Explicit construction for $D=5$ and numerical evolution}

In the following, we derive solutions of the stationary slice for
$D=5$ in terms of the variables of the BSSN formalism, and then,
demonstrate that the solutions are useful for checking numerical
relativity codes based on the BSSN formalism for $D=5$.

The fundamental variables in the BSSN formalism are different from
those in the so-called standard 3+1 formalism \cite{York}. In the
standard 3+1 formalism, the fundamental quantities are the intrinsic
metric $\gamma_{ij}$, the extrinsic curvature $K_{ij}$, the lapse
function $\alpha$, and the shift vector $\beta^i$ ($i,j=1,...,D-1$ in
$D$ dimensions). The initial values of these quantities are provided by
solving the constraint equations, and the subsequent evolution is
achieved by solving the evolution equations in certain coordinate
conditions (see e.g. Ref.~\cite{Smarr-York}).  The standard 3+1 
formalism prohibits a longterm stable numerical evolution, because
constraint violation modes grow in the presence of truncation 
error. In the BSSN formalism, the number of dynamical variables is
increased to suppress the source of such instability and to enable longterm
stable simulation.

The original form of the BSSN formalism was described in
Refs.~\cite{SN1995,BS1999}, and it was extended to for general
dimensionalities in Ref.~\cite{YS09}.  The definitions of dynamical
variables in the $D$-dimensional BSSN formalism are $\chi$: the
conformal factor; $\tilde{\gamma}_{ij}$: the conformal intrinsic
metric; $K$: the trace of the extrinsic curvature; $\tilde{A}_{ij}$: a
tracefree extrinsic curvature; and $\tilde{\Gamma}^i$: a auxiliary
$D-1$ variable.  $\chi$ and $\tilde{\gamma}_{ij}$ are defined by
\begin{equation}
\tilde{\gamma}_{ij}=\chi\gamma_{ij},
\end{equation}
where $\chi$ is determined so that the determinant 
of $\tilde{\gamma}_{ij}$ 
is equal to unity (note that we assume to use the Cartesian coordinates). 
The tracefree extrinsic curvature is defined by
\begin{equation}
\tilde{A}_{ij}:=\chi\left(K_{ij}-\frac{1}{D-1}\gamma_{ij}K\right),
\end{equation}
and $\tilde{\Gamma}^i$ is defined by
\begin{equation}
\tilde{\Gamma}^i:=-\frac{\partial\tilde{\gamma}^{ij}}{\partial x^j},
\end{equation}
where $\tilde{\gamma}^{ij}$ is the inverse of $\tilde{\gamma}_{ij}$, 
i.e., $\tilde{\gamma}^{ik}\tilde{\gamma}_{kj}=\delta^i_j$. 
The variables $\chi$, $K$, $\tilde{\gamma}_{ij}$, $\tilde{A}_{ij}$, and
$\tilde{\Gamma}^i$ are evolved, imposing gauge conditions for 
$\alpha$ and $\beta^i$.

For a numerical relativity simulation, the data of the stationary
slice are prepared in the Cartesian spatial coordinates
$(x,y,z,w)$. In the assumption that the intrinsic metric $\gamma_{ij}$
is conformally flat, $\tilde{\gamma}_{ij}=\delta_{ij}$
and $\tilde{\Gamma}^i=0$. Furthermore, the maximal slicing condition gives
$K=0$. Since the slice is spherically symmetric, it is sufficient to
prepare the data on the $x$-axis, i.e., $y=z=w=0$, on which the
following relations hold:
\begin{eqnarray} 
\beta^x&=&\beta^R, \qquad
\beta^{y} = \beta^{z}=\beta^{w}=0, \\
\tilde{A}_{yy}&=&\tilde{A}_{zz}=\tilde{A}_{ww}=-\frac{1}{3}\tilde{A}_{xx}
=-\frac{1}{3}\tilde{A}_{RR}, \\
\tilde{A}_{ij}&=&0~~~~~~~~{\rm for}~~i\neq j.
\end{eqnarray} 
Here, $R=\left(x^2+y^2+z^2+w^2\right)^{1/2}$ 
indicates the isotropic radial coordinate.
This implies that the data only for $\alpha$, $\beta^R$, $\tilde{A}_{RR}$, 
and $\chi$ are needed in the spherical polar coordinates. 

\begin{figure}[tb]
\centering
{
\includegraphics[width=0.45\textwidth]{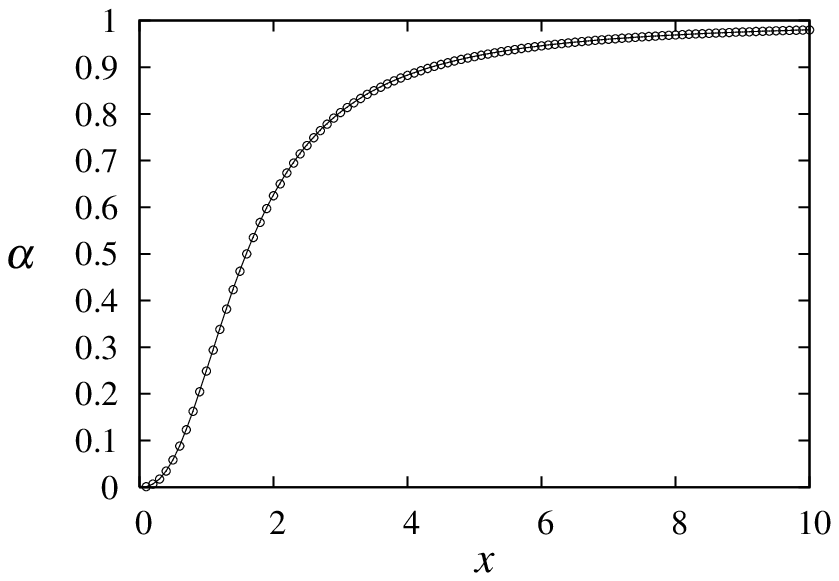}
\includegraphics[width=0.45\textwidth]{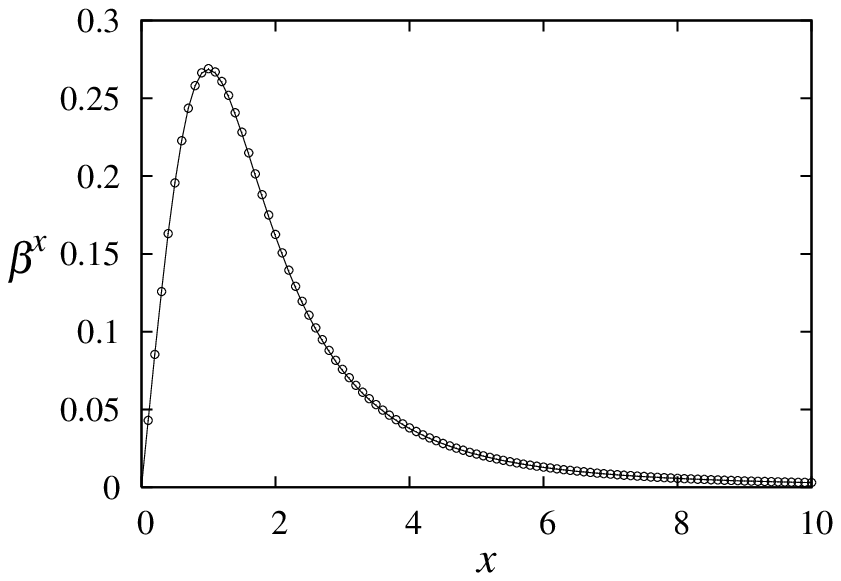}
\includegraphics[width=0.45\textwidth]{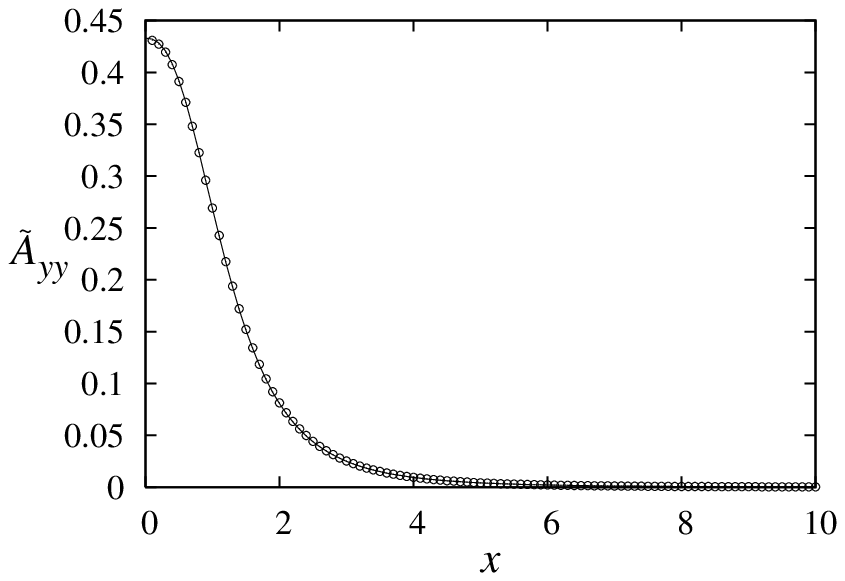}
\includegraphics[width=0.45\textwidth]{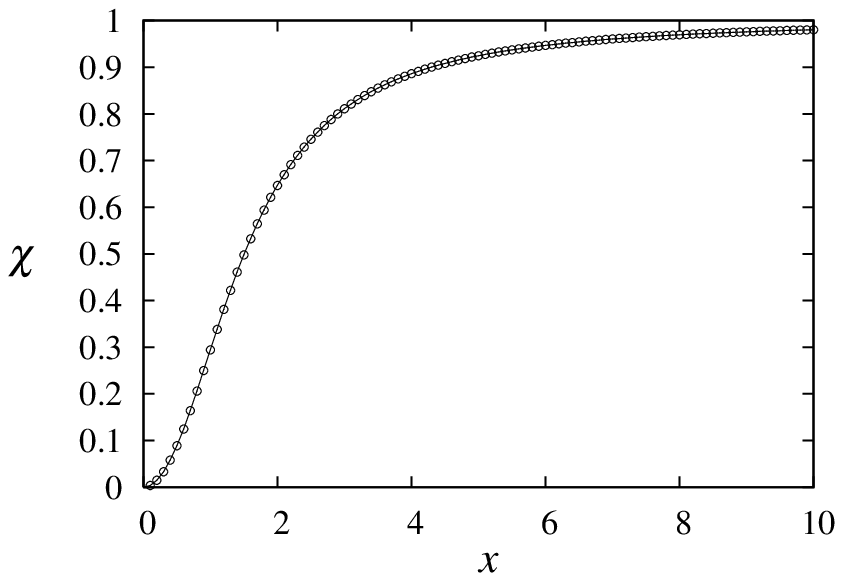}
}
\caption{The values of $\alpha$, $\beta^x$,
$\tilde{A}_{yy}$, and $\chi$ on the $x$-axis for the
stationary maximal hypersurface with $T=T_{\infty}$ (solid curves),
and the data after a longterm evolution by the time $t=100$ ($\odot$),
where the stationary maximal hypersurface was adopted as the initial data.  
The units of the length and time are $r_{\rm g}/2$. 
The data remain approximately stationary after the longterm evolution. 
}
\label{5D-case}
\end{figure}

For $D=5$, it is easy to perform integral in Eq.~\eqref{eq:I-def} to give 
\begin{equation}
R=\frac{r}{6}\left(3+\sqrt{3\left[(r_{\rm g}/r)^2+3\right]}\right)
\left(\frac{(5+2\sqrt{6})\left[3-2(r_{\rm g}/r)^2\right]}
{2(r_{\rm g}/r)^2+15+6\sqrt{2\left[(r_{\rm g}/r)^2+3\right]}}\right)^{1/\sqrt{6}}.
\label{eq:R-r-relation}
\end{equation}
The conformal factor $\chi$, the lapse function $\alpha$, and 
the $R$-component of the shift vector are given by
\begin{equation}
\chi=\psi^{-2}=\left({R}/{r}\right)^2, \label{eq:chi}
\end{equation}
\begin{equation}
\alpha=\sqrt{1-\left(\frac{r_{\rm g}}{r}\right)^2
+\frac{4}{27}\left(\frac{r_{\rm g}}{r}\right)^6},
\end{equation}
and
\begin{equation}
\beta^R=\frac{dR}{dr}\gamma^{-1}\beta
=\frac{2}{3\sqrt{3}}\frac{r_{\rm g}^3R}{r^4}.
\end{equation}
The $RR$-component of the tracefree extrinsic curvature is 
\begin{equation}
\tilde{A}_{RR}
=-\frac{2}{\sqrt{3}}\frac{r_{\rm g}^3}{r^4}.
\label{eq:tildeA}
\end{equation}
To describe the data in the isotropic radial coordinate $R$, we have
to give $r$ as a function of $R$.  Since the inversion of
Eq.~\eqref{eq:R-r-relation} cannot be done analytically, we
numerically derive the relation $r=r(R)$ and then generate the data as
functions of $R$.  The analytic initial values of $\alpha$, $\beta^x$,
$\tilde{A}_{yy}$, and $\chi$ on the $x$-axis are depicted by the solid
curves in Fig.~\ref{5D-case}.  Here, we adopt $r_{\rm g}/2$ as the
unit of the length. 


%
%

We evolve the initial data by using the numerical code recently
developed \cite{YS09} and show that the data indeed remains stationary
in the puncture gauge conditions.
We adopt a dynamical time slicing condition, 
\begin{equation}
\partial_t\alpha=-2\alpha K. \label{eq:1+log}
\end{equation}
This is a simplified version of the 1+log slicing condition~\cite{Bona_etal} 
that was studied in Ref.~\cite{Hannam_etal_1}. 
As the spatial gauge coordinates, 
we adopt the $\Gamma$-driver condition~\cite{Alcubierre:2002kk}
\begin{eqnarray}
\partial_t\beta^i= \frac{D-1}{2(D-2)}B^i \qquad {\rm and} \qquad
\partial_tB^i=\partial_t\tilde{\Gamma}^i-\eta B^i, \label{eq:G-driver}
\end{eqnarray}
where $\eta$ is constant, chosen to be $1/5r_g$--$20/r_g$. 

The stationary maximal slicing with the isotropic spatial coordinate
satisfies the coordinate conditions \eqref{eq:1+log} and
\eqref{eq:G-driver}, and thus, the numerical data have to be unchanged
during numerical evolution with these gauge conditions, if the initial
data for $\chi$, $\tilde{\gamma}_{ij}$, $K$, $\tilde{A}_{ij}$,
$\alpha$, and $\beta^i$ agree with those of the stationary maximal
hypersurface with $T=T_\infty$ covered by the isotropic coordinate
system, together with further initial data $B^i=0$.  Therefore, the
analytic solution for these variables given here is used for a test
simulation of numerical relativity codes based on the BSSN formalism
and puncture gauge. 

The data obtained by evolving the above analytic initial data with
$B^i=0$ is depicted by the circles $\odot$ in Fig.~\ref{5D-case}.  In
the numerical simulation, the outer boundary is located at a
sufficiently distant zone ($x=200$), and the grid size is uniformly
$\Delta x=0.1$. The value of $\eta$ is chosen for a wide range as
$\eta=1/(5r_{\rm g})$-- $20/r_{\rm g}$, and we confirm that the result
does not depend on the choice (but we found that numerical error
becomes large for very large values of $\eta$).  Figure~\ref{5D-case}
shows that the numerical data at $t=100$ agrees well with the initial
data $t=0$.  Thus, we conclude that the numerical data is
approximately stationary and unchanged in time.

%
\begin{figure}[tb]
\centering
{
\includegraphics[width=0.45\textwidth]{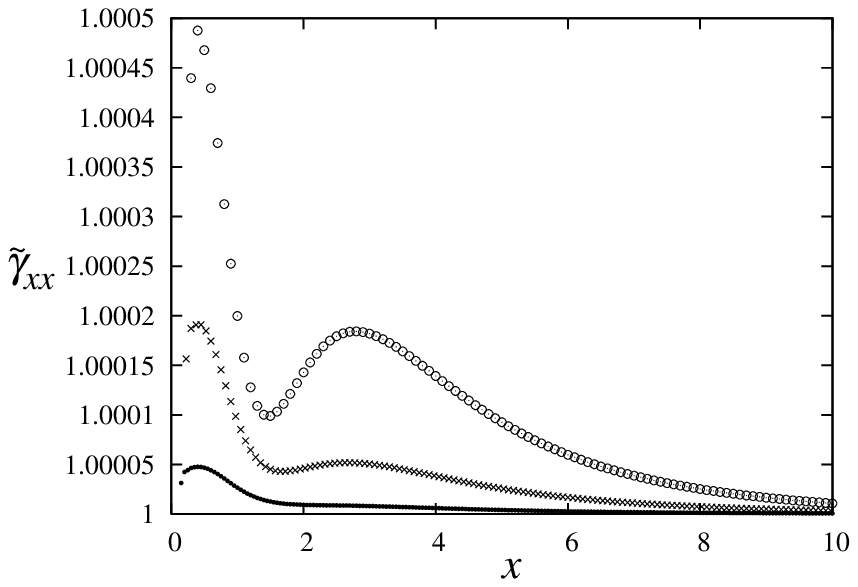}
\includegraphics[width=0.45\textwidth]{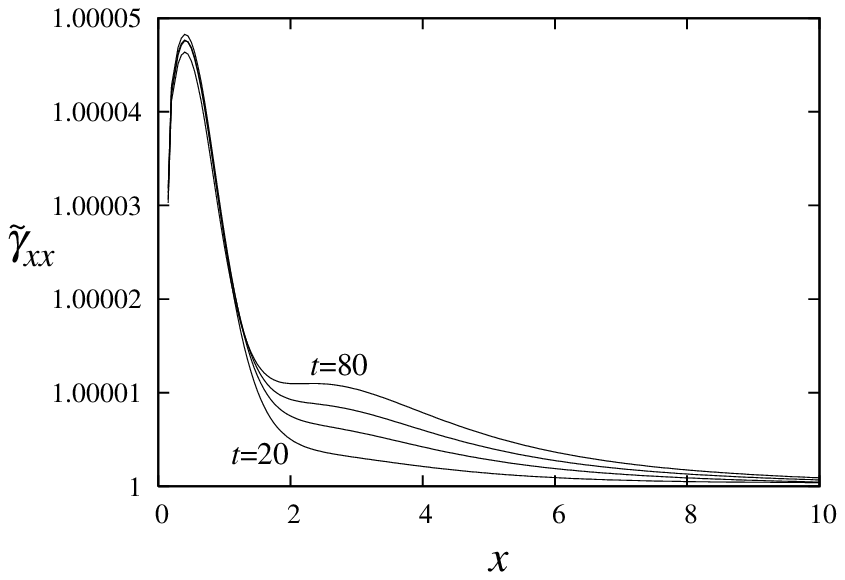}
}
\caption{Left panel: The value of $\tilde{\gamma}_{xx}$ at time $t=60$
for resolutions $\Delta x=0.1$ ($\odot$), $0.075$ ($\times$), and $0.05$
($\bullet$). The deviation from the analytic solution
$\tilde{\gamma}_{xx}^{(a)}=1$ becomes smaller as the resolution is
increased.  Right panel: Snapshots of $\tilde{\gamma}_{xx}$ for a
fixed grid spacing $\Delta x=0.05$ for time $t=20$, $40$, $60$, and
$80$. The error gradually increases during the evolution.  }
\label{numerical_error}
\end{figure}
%


The left panel of Fig.~\ref{numerical_error} shows the evolution of
numerical values of $\tilde{\gamma}_{xx}$ on the $x$-axis obtained by
$\eta=2/r_{\rm g}$ for several grid resolutions with $\Delta x=0.1$,
$0.075$, and $0.05$.  Because the analytic solution is
$\tilde{\gamma}_{xx}^{(a)}=1$, the deviation from unity indicates the
amount of the error. This figure clearly shows that the deviation is
decreased as the resolution is increased, and hence, the deviation
from the analytic data is caused only by the numerical error. Here, 
the spatial pattern of the error for a fixed time $t$ with $t\gg 10r_{\rm g}$ 
was found to depend on the grid resolutions. 
However, as the resolution is increased, 
the pattern becomes less dependent on the resolution and  
the error decreases approximately at the fourth order.
This implies that
our numerical solution achieves the four-order convergence
in the limit $\Delta x\to 0$ 
except at the region near the puncture 
(where the analyticity of solution is broken
and the numerical solution should not converge at the fourth order).
The right panel shows the snapshots of $\tilde{\gamma}_{xx}$ for
$t=20$--$80$ for the grid size $\Delta x=0.05$. 
Although the error grows during the evolution, the
growth rate is small.  These results illustrate both the accuracy of
our numerical simulations and the effectiveness of the foliation by
the stationary maximal hypersurfaces as a benchmark for checking a
five-dimensional numerical relativity code.

\section{Summary}

We have studied the foliation of the $D$-dimensional
Schwarzschild-Tangherlini spacetime by the two kinds of one-parameter
family of maximal hypersurfaces: the reflection symmetric foliation
with respect to the wormhole slot and the stationary foliation.  We
have shown that the both foliations have the singularity avoidance
nature for $D\geq 5$, as in the case of $D=4$.  It is also shown that
each hypersurface of the stationary foliation has an infinitely long
trumpet-like shape in the neighborhood of the black-hole puncture
located at the origin of the isotropic coordinate.  Because the
stationary foliation is the attractor of the numerical evolution by a
dynamical slicing condition \cite{Hannam_etal_1}, both the maximal
slicing condition and the dynamical slicing conditions will have the
preferable nature for the puncture method for $D\ge 4$. We presented
the explicit solution of the stationary foliation for $D=5$, and
showed, by performing the numerical simulation, that it is useful for
a benchmark test of $D=5$ numerical relativity codes.

The remaining issues to be explored are as follows.  Although we
expect that the puncture gauge condition based on a dynamical slicing
and $\Gamma$-driver gauge condition also works well for many issues
in the higher-dimensional numerical relativity, as demonstrated in
Ref.~\cite{Hannam:2006vv} in the four-dimensional case, more detailed
studies for a variety of spacetimes are obviously needed.  For
example, it is important to figure out the gauge conditions suitable
for simulating higher-dimensional rotating black hole spacetimes
(i.e. Myers-Perry black holes \cite{MP86}) and for simulating black
holes with high velocity. These are issues to be studied.

\acknowledgments

The authors would like to thank colleagues in the theoretical
astrophysics and gravity group in Osaka City University. HY is
supported by JSPS (program of Postdoctoral Fellow for Research
Abroad).  MS was in part supported by Grant-in-Aid for Scientific
Research (21340051) and by Grant-in-Aid for Scientific Research on
Innovative Area (20105004) of the Japanese Monbukagakusho.

\appendix

\section{Kruskal extension and embedding of maximal hypersurfaces}

In this section, we analyze the Kruskal extension
of the $D$-dimensional Schwarzschild-Tangerlini spacetime
and explain how to embed maximal hypersurfaces in it. 
We start from the standard metric \eqref{eq:standard} in the
static coordinates $(\tau, r)$ and consider the region $r > r_{\rm g}$. 
First, we introduce the so-called tortoise coordinate $r_*$,
\begin{equation}
r_*:=\int\frac{dr}{1-\left(r_{\rm g}/r\right)^{D-3}}
=r+\frac{r_{\rm g}}{D-3}\left[\ln\left|\frac{r}{r_{\rm g}}-1\right|+G(r)\right].
\end{equation}
Here, defining 
\begin{multline}
G_n(r):= \cos\left(\frac{2n\pi}{D-3}\right)
\ln\left|\left(\frac{r}{r_{\rm g}}\right)^2-\frac{2r}{r_{\rm g}}
\cos\left(\frac{2n\pi}{D-3}\right)+1\right| 
\\
+2\sin\left(\frac{2n\pi}{D-3}\right)
{\arctan} \left(\frac{\cos[2n\pi/(D-3)]-r/r_{\rm g}}{\sin[2n\pi/(D-3)]}\right),
\end{multline}
the function $G(r)$ is given by
\begin{equation}
G(r)=\sum_{n=1}^{(D-4)/2}G_n(r), 
\qquad \textrm{for even}~D\ge 4,
\end{equation}
and 
\begin{equation}
G(r)=\ln\left|\frac{r}{r_{\rm g}}+1\right|^{-1}+\sum_{n=1}^{(D-5)/2}G_n(r),
\qquad \textrm{for odd}~D\ge 5.
\end{equation}
It is easily seen that $G(r)$ is regular for $r \geq 0$. 
Then, we introduce the Kruskal null coordinates as
\begin{eqnarray}
U&=&-r_{\rm g}\exp\left[-(D-3)(\tau-r_*)/2r_{\rm g}\right], 
\label{def:coordinate-U}\\
V&=&+r_{\rm g}\exp\left[+(D-3)(\tau+r_*)/2r_{\rm g}\right].
\label{def:coordinate-V}
\end{eqnarray}
In these coordinates, the metric \eqref{eq:standard}
is reduced to the following form:
\begin{equation}
ds^2=-\left(\frac{2}{D-3}\right)^2e^{-(D-3)r/r_{\rm g}-G(r)}
\sum_{n=1}^{D-3}\left(\frac{r_{\rm g}}{r}\right)^n
dUdV+r^2d\Omega_{D-2}^2.
\label{metric:Kruskal}
\end{equation}

Now, we can extend the spacetime
in a similar manner to the four-dimensional case.
The coordinates $U$ and $V$ introduced by Eqs.~\eqref{def:coordinate-U}
and \eqref{def:coordinate-V} are restricted to the region $U<0$
and $V>0$. However, since the metric is regular at $U=0$ and $V=0$,
the spacetime is extended to the region $U>0$ or $V<0$.
The maximally extended spacetime consists of four regions,
and the three regions obtained by the extension are: 
the black hole region $U>0$ and $V>0$,
the white hole region $U<0$ and $V<0$, and 
the other region $U>0$ and $V<0$ outside of the two holes 
beyond the wormhole slot. 
The relation between the coordinates $(U,V)$ and the static
coordinates $(t,r)$ in each extended region 
is given by appropriately changing the sign of 
Eqs.~\eqref{def:coordinate-U} and \eqref{def:coordinate-V}.
Note that the range of $r$ is $0<r<r_{\rm g}$ 
in the black and white hole regions 
while $r_{\rm g}<r$ in the two outside regions. 
The line element \eqref{metric:Kruskal} 
is regular everywhere except at the two physical curvature singularities,
$r=0$, in the black and white holes. 
See Fig.~\ref{fg:stationary} for the structure of the 
maximally extended spacetime.

In order to embed maximal hypersurfaces in the Kruskal
diagram, we consider the coordinate transformation
between the Kruskal coordinates $(U,V)$ and 
the coordinates $(t,r)$ for maximal slicing.  
Since $(U,V)$ and $(\tau,r)$ are related through
Eqs.~\eqref{def:coordinate-U} and \eqref{def:coordinate-V} 
whereas $(\tau,r)$ and $(t,r)$ are related through Eqs.~\eqref{eq:dtaudt}
and \eqref{eq:dtaudr}, we obtain
\begin{eqnarray}
\left.
\frac{\partial U}{\partial r}\right|_t
&=&\frac{D-3}{2r_{\rm g}}
\left[1-\left(\frac{r_{\rm g}}{r}\right)^{D-3}\right]^{-1}
\left(1+\gamma^{1/2}Tr^{-(D-2)}\right)U, \label{eq:dUdr}\\
\left.
\frac{\partial V}{\partial r}\right|_t
&=&\frac{D-3}{2r_{\rm g}}
\left[1-\left(\frac{r_{\rm g}}{r}\right)^{D-3}\right]^{-1}
\left(1-\gamma^{1/2}Tr^{-(D-2)}\right)V. \label{eq:dVdr}
\end{eqnarray} 
Once the arbitrary function of integration $T(t)$ is determined, 
we obtain a maximal hypersurface in the Kruskal diagram by  
integrating these two equations.



\end{document}